\def\MYJOURNAL{0} %%PR%

\if 0\MYJOURNAL%
\documentclass[3p, preprint, sort&compress, a4paper]{elsarticle}
\makeatletter
\def\ps@pprintTitle{%
  \let\@oddhead\@empty
  \let\@evenhead\@empty
  \def\@oddfoot{\reset@font\hfil\thepage\hfil}
  \let\@evenfoot\@oddfoot
}
\makeatother
\else \if 1\MYJOURNAL%
\documentclass[journal,10pt,twocolumns]{IEEEtran}
\usepackage[numbers,sort&compress]{natbib}
\fi
\fi

\usepackage{graphicx}
\usepackage{subfig}
\usepackage{amsmath, amsfonts, amssymb, bm}
\usepackage{booktabs}
\usepackage{geometry}%
\usepackage{enumitem}

\usepackage{fourier}
\DeclareMathAlphabet{\mathcal}{OT1}{pzc}{m}{it}
\DeclareSymbolFont{letters}{OML}{cmm}{m}{it}

\graphicspath{{./Graphics/}}

\usepackage{color, xcolor, colortbl}
\usepackage{hyperref}
\hypersetup{
  colorlinks,
  citecolor=blue,
  filecolor=blue,
  linkcolor=blue,
  urlcolor=blue
}

\usepackage{amsmath, bm, accents}
\usepackage{amssymb}
\usepackage{graphicx}
\usepackage{algorithmic}
\usepackage{algorithm}
\usepackage{subfig}
\usepackage{color}
\usepackage{amsmath}
\usepackage{rotating}%

\newdefinition{rmk}{Remark}

\begin{document}

\title{Assessment of multiple-biomarker classifiers:\\Fundamental principles and a proposed
  strategy}

\author[WAY]{Waleed~A.~Yousef\corref{cor1}}
\ead{wyousef@GWU.edu, wyousef@fci.helwan.edu.eg}
\cortext[cor1]{Corresponding Author}

\address[WAY]{Ph.D., Computer Science Department, Faculty of Computers and Information, Helwan
  University, Egypt.\\ Human Computer Interaction Laboratory (HCI Lab.), Egypt.}

\begin{abstract}
  The multiple-biomarker classifier problem and its assessment are reviewed against the background
  of some fundamental principles from the field of statistical pattern recognition, machine
  learning, or the recently so-called ``data science''. A narrow reading of that literature has led
  many authors to neglect the contribution to the total uncertainty of performance assessment from
  the finite training sample. Yet the latter is a fundamental indicator of the stability of a
  classifier; thus its neglect may be contributing to the problematic status of many studies. A
  three-level strategy is proposed for moving forward in this field. The lowest level is that of
  construction, where candidate features are selected and the choice of classifier architecture is
  made. At that point, the effective dimensionality of the classifier is estimated and used to size
  the next level of analysis, a pilot study on previously unseen cases. The total (training and
  testing) uncertainty resulting from the pilot study is, in turn, used to size the highest level of
  analysis, a pivotal study with a target level of uncertainty. Some resources available in the
  literature for implementing this approach are reviewed. Although the concepts explained in the
  present article may be fundamental and straightforward for many researchers in the machine
  learning community they are subtle for many practitioners, for whom we provided a general advice
  for the best practice in \cite{Shi2010MAQCII} and elaborate here in the present paper.
\end{abstract}%


\maketitle

\section{BACKGROUND}
Multidimensional models and multivariate analysis are becoming ubiquitous in contemporary bioscience
and the associated fields of medical diagnostics and therapeutics and the analysis is
correspondingly complex \citep{Butcher2004SysBioDrug}. On one quite visible front, progress started
toward establishing a framework for obtaining reliable and reproducible measurements from
high-dimensional gene-expression microarrays in a clinical or regulatory setting, e.g., the report
from the first phase of the Microarray Quality Control Consortium (MAQCI) \citep{Shi2006MAQCproj}
and the subsequent distinctions and discussions generated by it
\citep{Klebanov2007StatMethodMicroArr, Liang2007MAQCpapersCracks}.

\bigskip

One of the natural next steps is that of training and testing a predictive model to use such
multidimensional, or multiple-biomarker, data to perform a task of binary classification, e.g., to
discriminate between the presence or absence of a specified disease state, or between an aggressive
and an indolent sub-class of the disease, or to discriminate prospectively between responders and
non-responders to a specified therapy, among many other applications. A generalization of this task
from binary to continuous outcomes is prediction of patient survival times, conditional on a
specified treatment. The most commonly studied disease in the multiple-biomarker field today is
cancer. The second phase of the MAQC consortium (MAQCII), had the goal of consensus development on
the design and validation of such predictive models. This is a challenging undertaking. The task of
building such models is not generally well-posed; moreover, a coherent consensus on the appropriate
methodology for assessing a resulting model, here a multiple-biomarker classifier, has yet to
emerge. We reported our main conclusion of the MAQCII project and guides for the best practices to
practitioners in our Nature Biotechnology paper \cite{Shi2010MAQCII}, along with more detailed
analysis in \cite{Chen2012UncertEst}.

The purpose of the present paper is to elaborate on these two publications; an earlier attempt was
our publication \cite{Wagner2008FiniteTrainingRadSLM}. However, for publications size limitation not
all of the concepts were treated fairly. The present paper thus tries to contribute to the consensus
development process in this very broad and challenging field by focusing on some of the more subtle
issues that are usually ignored by practitioners, and offering a candidate framework for the task of
assessing a multiple-biomarker classifier that takes those issues into account.

\bigskip

It is well known and widely remarked that the multiple-biomarker problem---particularly in the very
high-dimensional formats of contemporary gene expression measurements---is an ill-conditioned
problem at the outset. To wit: measurement of tens of thousands of biomarkers are made in parallel
on (typically) just a few hundred patients; thus there is insufficient data in typical datasets to
directly determine a solution to a specified diagnostic task that is close to a good solution in the
population. There is no general purpose approach to attacking such ill-posed problems. However, a
great variety of tools are being developed (and many are reviewed in \cite{Parmigiani2003AnaGene,
  Speed2003StatAnaGene, McLachlan2004DiscAna, Simon2003DesignAnaDNA}) to reduce the dimensionality
of the biomarker or feature space to a point where a manageable set of informative features has a
chance to be reliably observed and combined using a statistical learning machine, most simply and
thus most frequently a binary or two-class classifier. Some form of ``internal validation'' usually
accompanies this exercise; this must be followed up, of course, by a form of ``independent external
validation.'' We will here propose a more general variation on that approach to assessment of the
resulting classifier based on fundamental concepts, and contrast this with contemporary experience
with the less general approach.

\section{STATISTICAL LEARNING MACHINES}
The task that we address in this manuscript is the assessment of the performance of a binary
classifier developed in the context sketched above. Such a classifier is a member of the broad class
of ``statistical learning machines'' (SLM). For a full account of SLM the reader is referred to the
seminal text \cite{Hastie2001TheElements}, or for a very concise summary to
\cite{Yousef2008SLM-arxiv}. The classical paradigm in that field involves collecting data in two
stages: first, the so-called ``training'' or ``design'' sample, from which to learn the statistical
structure of the problem and design the classifier appropriately; and second, the so-called
``testing'' sample to assess the performance of the designed classifier. A central premise of the
present manuscript is that the finite size of both the training and the testing sample sets
contributes to the variance of interest to us when we estimate the performance of a classifier and
estimate the uncertainty in the performance estimate. Our arguments follow below after further
historical motivation.

\subsection{Finite training sample analyses}
\begin{table}[t]\centering
  \begin{tabular}{lp{3in}l}\toprule
    \bfseries Classifier & \bfseries VC dimension & \bfseries Reference\\
    \midrule
    Hyperplane in $d$ dimensions & $d+1$ & Theorem of Cover \citep{Cover1965GeomStatProp} \\
    Polynomial surface & number of parameters & \cite{Cover1965GeomStatProp}\\
    Artificial Neural Networks (ANN) & $w<d_{VC}<2w\log_{2}(e\ n)$, where $w$ is the no. of weights and $n$ is the number of nodes (a ``node'' is a threshold unit in a neural network) & \cite{Baum1989WhatSize}\\
    Support Vector Machine (SVM) & number of support vectors & \cite{Vapnik2000Nature}\\
    $k-$NN & $\sim 1/k$ & \cite{Hastie2001TheElements}\\
    \bottomrule
  \end{tabular}
  \caption{VC dimension, $d_{VC}$, for some common classifier architectures.}\label{TabVCdim}
\end{table}%
An analysis of the performance of the binary classifier in the context of finite data, for the case
where the data is distributed as multivariate normal, goes back to the 90's and was given by
Fukunaga in the second edition of his classic text \cite{Fukunaga1990Introduction} (see also
\cite{Fukunaga1989EstimationOf}). He develops the lowest-order finite-sample analysis for both the
bias and the variance of the performance of the so-called Bayes classifier for that problem under
the condition of finite training and finite testing samples, respectively $N_{train}$ and
$N_{test}$. The Bayes classifier is constructed from the relative probability of the data under the
two classes for the problem, i.e., the likelihood ratio. It is shown in
\cite{Fukunaga1990Introduction}, among many others, that this classifier is optimal in terms of
several fundamental criteria in the field of statistical decision making. In practice, however, the
true probability distributions for the classes will not be known; even if they could be known, the
parameters of the distributions would have to be estimated from the data, so the true likelihood
ratio would not be known. (Although it is a bit off topic, the connection between the multinormal
distribution of the features and the normal distribution of the scores generated by the classifier
is elaborated in \cite{Yousef2019PrudenceWhenAssumingNormality-arxiv}.)

\bigskip

The ``bias'' analyzed by Fukunaga in this problem is the difference between the mean performance
when the classifier is trained with an infinite sample of trainers and the mean performance when the
classifier is trained with a specified finite number of training cases. This bias is therefore just
another way of speaking of being on the knee or shoulder of a learning curve---short of reaching the
infinite-training asymptote \citep{Hastie2001TheElements}. It is not a bias in the sense of authors
who attempt to estimate the actual performance of a classifier conditioned on a single given finite
dataset \citep{Efron1983EstimatingTheError, Efron1997ImprovementsOnCross}. Here we will not further
refer to the concept of bias in the sense used by Fukunaga, but we note that this is the central
quantity of the paper on microarray sample size planning by Dobbin and Simon
\citep{Dobbin2007SampleSizePlannig}, to which we return below.

\bigskip

Fukunaga next analyzes the variance of estimates of classifier performance given the limitation of
finite training and testing. The figure of merit used by Fukunaga is total error or probability of
misclassification (PMC). The preferred figures of merit in our present community of interest are
based on the receiver operating characteristic (ROC) curve, i.e., the graph of the true-positive
fraction of classifications (or sensitivity) vs. the false-positive fraction (or one minus the
specificity). Common summary measures include the area under the ROC curve, or the partial area
under a specified portion of the curve \citep{Walter2005ThePartialArea, Jiang1996AReceiverOperating,
  Dodd2003PAUCEstReg, Yousef2013PAUC}, or---usually with less statistical power---a particular point
in ROC space (see \cite{Wagner2007AssMedImgTutorial} for a contemporary tutorial review). In any
case, the effects to be discussed below may differ quantitatively depending on the figure of merit,
but so far in our analysis and simulations they have been found to be qualitatively similar.

\bigskip

In Fukunaga's analysis of the variance of the performance estimate from a finite number of trainers,
$N_{train}$, and an infinite number of testers, the lowest-order term, namely, 1/$N_{train}$, vanishes
as a result of doing the analysis in the neighborhood of the Bayes classifier. The lowest-order
surviving term is then quadratic in that small quantity, i.e., proportional to $(1/N_{train})^2$, and
considered negligible compared to the expected $1/{N_{test}}$ dependence for finite testers. This
outcome has led to the conventional wisdom among students of that excellent text that the variance
of performance estimates in the context of finite trainers and finite testers is ``dominated by the
finite test sample'' \citep[][p. 218]{Fukunaga1990Introduction}. This situation is one possible
explanation for the relative lack of attention to the finite-training sample in the
literature. Another reason is that many researchers in the field of statistical learning work in
data-rich applications, e.g., identification of hand-written numerals for zip-code recognition, iris
recognition for security applications, and satellite imagery, among others.

\bigskip

A point of departure for the present paper is that this result was derived for a special category of
classifier, in particular one operating in the neighborhood of the Bayes classifier and under the
multivariate normal assumption. However, one cannot expect to be in that neighborhood when working
with other classifiers and with high-dimensional datasets whose distributions are unknown---so it is
unwise to follow that conventional wisdom. Moreover, we have found using a mixed model for the
components of variance that---for the problem where competing classifiers are being compared
\citep{Beiden2003AGeneralModel}---the finite-training sample contribution to the variance dominates
that from the finite test sample. Finally, further analysis\footnote{B.D. Gallas, Medical Image
  Perception Conference X, Duke University, Durham NC, Sept. 11-14, 2003} shows that termination of
the analysis in the manner of Fukunaga neglects important cross terms of the form
$1/(N_{train}N_{test})$. We conclude that the finite-training sample contribution to the total
variance is not a negligible quantity. We will augment this fundamental statistical argument below
with further arguments based on current practice.

\bigskip

For the lowest-order (and simplest) classifiers represented by the linear discriminant, the
finite-training sample effects increase with the dimensionality $d$ of the feature space (here, the
number of genes or biomarkers included in the classifier) \citep{Fukunaga1990Introduction,
  Fukunaga1989EstimationOf}, \cite[][e.g., Eq. 2.28]{Hastie2001TheElements},
\cite{Chan1999ClassifierDesign, Yousef2006DSc, Yousef2019PrudenceWhenAssumingNormality-arxiv}. For
more complex classifiers, the effects can be complicated to model. We simply assume here from
simulations that the effects increase with dimensionality and/or complexity, but not in a way that
is straightforward to formulate. Nevertheless, there may be a semi-quantitative approach to
capturing the trend, as we discuss next.

\bigskip

A concept that is used to characterize the ``complexity'' of a classifier is that of classifier
``capacity''. The simplest example of capacity is provided by a theorem whose history and proofs
have been given by Cover \citep{Cover1965GeomStatProp} (see also Ripley
\citep{Ripley1996PRandNN}). In effect---and in our present context---this theorem says:

Almost every possible labeling of multiple-biomarker samples in two classes will be perfectly
separable by a simple hyperplane, unless the number of samples is a multiple greater than unity
(when only a few features are used) or about two (when roughly 25 or more features are used) times
the number of dimensions (features or biomarkers)---unless the samples lie in a lower dimensional
subspace of the original one. In the latter case, the problem is analyzed in the same way in the
lower-dimensional space.

\bigskip

The number of randomly labeled patterns a learning machine can store and perfectly separate is
referred to as the ``capacity'' of the learning machine. For a hyperplane, then, this is $d + 1$ in
low dimensions or approximately $2(d + 1)$ in higher dimensions (e.g., $> 25$). Until the number of
available patients well exceeds this capacity, one cannot expect a classifier trained with these
patients to ``generalize'' to previously unseen ones \citep{Duda2001PatternClassification}, i.e., to
make a transition from the ill-conditioned or ``under determined'' domain to an ``over determined''
one, the goal of the investigation.

\bigskip

The concept of capacity is closely related to a measure of classifier complexity referred to as VC
dimensionality (after Vapnik and Chervonenkis). A concise instructive review of this topic is given
by Bishop \citep{Bishop1995NeurNet} and also by Hastie et
al. \cite[][Ch. 7]{Hastie2001TheElements}. The VC dimension, $d_{VC}$, is a measure of the intrinsic
storage capacity of a learning machine for random patterns, in the spirit of this concept as used by
Cover in the reference cited above. In Vapnik's deep theory of statistical learning machines
\citep{Vapnik2000Nature, Vapnik1998StatLerningTh}, probabilistic bounds on the optimism of the error
observed in the training sample are given in terms of a nonlinear function of the VC dimension of
the class of functions used by the learning machine. Qualitatively, the optimism increases with
$d_{VC}$ and decreases with the size of the training sample $N_{train}$. The $d_{VC}$ has been
calculated or approximated for only a few classes of learning machines. Some examples are provided
in Table \ref{TabVCdim} (references are given in the table).

\bigskip

The $d_{VC}$ concept is of interest to us in the present paper since it may be useful to developers
of learning machines to think of their sample size in units of that dimension. A popular rule of
thumb (compare Fukunaga and Hayes \cite{Fukunaga1990Introduction, Fukunaga1989EstimationOf} on the
linear classifier) is that the training sample size should be some multiple of the dimensionality of
the feature space to constrain the optimism. Any such multiple will depend on the classifier
complexity and the intrinsic class separability. Vapnik refers to the ``small sample'' regime in his
research when the ratio of the number of training patterns to the $d_{VC}$ is smaller than about
20. It is remarkable that Vapnik's analysis of this regime includes no measure of the actual or
estimated separability of the data; this may explain in part the observation that the bounds
provided by the VC dimension are often very loose \cite[][Ch. 7]{Hastie2001TheElements} or,
equivalently, conservative. For example, if there is great separability of the data, Vapnik's
analysis is clearly too demanding of samples. The factor of two in the discussion of
\cite{Duda2001PatternClassification} and the factor of twenty in the analysis of Vapnik
\citep{Vapnik2000Nature} may be among the reasons for the popular impression that there is some
practical multiple or safety factor in the order of 5-10---but no general result
exists. Nevertheless, the concept of the VC dimension remains an important point of reference for
thinking in terms of trends on the requirements for a training set size not in absolute but in
relative terms \cite[][Ch. 7]{Hastie2001TheElements}.

\bigskip

More intuitive or tangible measures of classifier complexity are those of the effective number of
degrees of freedom of the model in the classifier or the effective number of parameters being
fit. \cite{Hastie2001TheElements} demonstrates how to estimate these for linear models as well as
some nonlinear ones.

\bigskip

One of the first major publications on gene-expression-based classifiers was that of Golub and
colleagues on classification of human acute leukemias \citep{GolubEtal1999MolecClass}. We consider
their work in the light of the concepts above. Those authors identified 50 ``informative genes''
among many thousands of candidates. They then trained a large number of classifiers for the task of
discriminating acute lymphoblastic leukemia (ALL) from acute myeloid leukemia (AML) based on the
levels of expression of these genes in patients' bone marrow samples, and found success in the
training set almost independent of a wide range of choice among those genes. A peculiar feature of
this work is that only 38 patients were in the training set. But the theorem of Cover tells us that
almost every possible labeling of multiple-biomarker samples in two classes (here, ALL versus AML)
will be perfectly separable by a simple hyperplane, unless the size of the sample is a multiple
greater than about two (for such high-dimensional data) times the number of features (or biomarkers)
used. In that light, almost perfect separability of the two classes of those samples would be
expected---unless those genes dwell on a much lower dimensional subspace. Golub et al. reported that
``Predictors based on between 10 and 200 genes were all found to be 100\% accurate\ldots'' This
discussion raises the possibility that some of the results are purely a consequence of a small
sample size in high-dimensional geometry and the theorem of Cover in that context
\cite{Cover1965GeomStatProp}.

\bigskip

On the other hand, inspection of the levels of expression of the 50 genes and the two classes in the
paper by Golub et al. makes it visually obvious that those candidate genes are strongly
co-expressed---so we cannot invoke the theorem of Cover until the dimensionality of their
appropriate subspace is determined. This would require a principal components or Singular Value
Decomposition (SVD) analysis \cite[][Ch. 14]{Hastie2001TheElements}, which the authors did not
provide. Nevertheless, the authors checked their classifiers against an independent test set of 34
cases, with very good results. It is interesting that their total of 72 cases is still smaller than
twice the dimension of their typical feature space (all the more so, for their larger spaces)---the
minimum required by naive application of the theorem of Cover cited above
\citep{Cover1965GeomStatProp}. In their favor, it seems very likely that their genes lie on a much
lower dimensional subspace, and very unlikely given their level of internal and external validation
that all of their results are a geometrical coincidence. Nevertheless, in the view of us, a measure
of the training sample size in units of some measure of complexity, effective dimensionality, or
degrees of freedom of a classifier should be a critical issue for such experimenters to address and
provide to their readers. It is uncommon to see this in reported studies.

\bigskip

This discussion and the underlying literature cited are some highlights of what can be achieved from
formal analysis of the training problem. In practice, the issue becomes an empirical one, which we
now address.

\section{WHAT PERFORMANCE TO ESTIMATE}
\subsection{Which point estimate?}\label{sec:which-point-estimate}
What is the quantity of interest to us? Are we interested in the performance of the classifier
conditional on a given available finite-sample training set? Or are we interested in the
unconditional performance, i.e., the mean over the population of training sets of that same finite
size? We adopt the latter point of view. The unconditional performance is clearly more
representative of the technology being evaluated. Further arguments for this will be contained in
the discussion presented below for what variance to estimate. A more subtle point, however, when
estimating the performance using resampling from only one available dataset
(Sec.~\ref{sec:level-one-or}) is the following. Estimates of the conditional performance based on
any of the commonly used versions of cross validation do not correlate well with the population
result (i.e., corresponding to an infinite number of testers), on a sample-by-sample basis
\citep{Efron1997ImprovementsOnCross, Yousef2006DSc, Zhang1995AssessingPrediction}. In the light of
this remarkable result, there is no basis for preferring the conditional performance to the mean
over the population of training sets of the same finite size. Moreover, \textit{``... we would
  guess, for any other estimate of conditional prediction error''}
\citep[Sec. 7.12,][]{Hastie2009ElemStat}. We elaborate more on these concepts in our work
\cite{Yousef2019AUCSmoothness-arxiv, Yousef2019LeisurelyLookVersionsVariants-arxiv}.

\subsection{What variance to estimate?}
The next natural question regards the variance estimate. Should one estimate the uncertainty or
variability that one would observe if the classifier is fixed, and on replication of the experiment
only a new random sample of testers is drawn? Or should one estimate the variability that one would
observe if, on replication of the experiment, a new random sample of trainers is drawn as well as a
new random sample of testers? Of course, confidence intervals for the latter approach will in
general be wider than those for the former approach.

% The point of view of the present authors is that the second, or more general, uncertainty should
% be specified.
Classifier designers sometimes argue that their training is fixed or ``frozen.'' But if one intends
to take a particular technology seriously, it is to be expected that classifier designers will
continue to build their database and update their algorithms in order to move up the learning
curve---unless they have some demonstration that they have reached the expected asymptotic
performance or maturity of their technology. In our practical experience, however, no investigator
of a multiple-biomarker predictive test has yet demonstrated the existence of a mature technology
and so it is unreasonable to consider any algorithm as really frozen. It will thus evolve and its
performance can then migrate well outside the confidence intervals estimated only on the basis of
the finite test sample. This is a principal motivation for the more general approach.

Moreover, one can argue that performance conditional on any particular training set does not
characterize the diagnostic technology under evaluation. That technology is neither the set of
biomarkers selected nor the structure of the classifier used to fuse them, but rather the
combination of the biomarkers and the classifier. The particular training sample that is drawn is a
random effect in the overall process. Neglecting that randomness in the assessment would mean that
something that is not technologically or scientifically well-defined is being assessed.

\bigskip

Further motivation for our position can be drawn from an exchange in the multiple-biomarker
literature. S.S. Dave and colleagues \citep{Dave2004PredOfSurv} trained a model for
prediction-of-survival on 95 biopsy specimens from patients with untreated follicular lymphoma. They
tested the model by validating it on an independent test set of 96 specimens from the same
population, and found a highly significant association of their model with survival in the
independent test set $(p<0.001)$.

Tibshirani acknowledged the authors' result that their recipe gave a significant $p$-value when used
on the test set, but went on to point out that ``the result is extremely fragile''
\citep{Tibshirani2005ImmSigLymphoma}. When the training and test sets are swapped, and the authors'
recipe for training is applied, the authors' model does not emerge. Additional analyses ``suggest
that their result occurred by chance and is not robust or reproducible.''

\bigskip

The robustness or reproducibility (versus fragility or irreproducibility) of a classifier are
qualitative ways of thinking of what we have called the ``intrinsic variability of a classifier
trained with a finite sample'' \citep{Yousef2006AssessClass}. A quantitative discussion of these
issues would require a number of formal definitions, and is beyond our present scope. Details can be
found, however, in \cite{Yousef2006AssessClass, Chen2012ClassVar}.

The Tibshirani-Dave exchange leads us to conclude that even what we refer to as ``the traditional
hygiene of independent training and test sets'' may, alone, fail to probe the robustness of a model
and should give us pause regarding our level of satisfaction with that as the ultimate test.

And finally, our position on estimating the total variability from the finite training set as well
as the finite testing set is consistent with our position that one is ultimately interested in the
unconditional performance, not the performance conditioned on a particular training set, when
assessing such a diagnostic technology.

Additional support of our position comes from considering the limitations of cross validation, whose
many forms are among those typically used by investigators for their internal validation procedure.

\subsection{Cross validation: caveats}
Many authors feel comfortable using Cross Validation (CV) training and testing of a classifier and
some measure of variability resulting from that exercise. So-called $K$-fold CV (CVK) involves
partitioning the available data into $K$ subsets; on each of $K$ passes (or ``folds'') a different
one of the subsets is held out for testing and the remaining $K-1$ are combined into the training
set. The limiting case is the very popular Leave-One-Out or ``round robin'' training-and-testing CV
(LOOCV)---an approach that is widely accepted in the pattern recognition community as a method for
reducing the pessimistic bias (in the sense of \cite{Fukunaga1990Introduction,
  Fukunaga1989EstimationOf}, and \cite[][P. 215]{Hastie2001TheElements}) in classifier performance
associated with a finite training set size. Many other versions and variants exist for CV; however,
many of them my be redundant as explained in
\cite{Yousef2019LeisurelyLookVersionsVariants-arxiv}. Although the CV approach can be very useful
for ``tuning'' or optimizing the parameters of a model, and for estimating the performance, it has
some theoretical limitations. In addition to the conditional/unconditional performance paradox discussed in Section~\ref{sec:which-point-estimate} there is another theoretical issue for discussion. 
% this approach does not provide a measure of robustness of the model. The training sets
% are very similar throughout the cross validation. There is thus no estimation of the uncertainty
% from the finite training sample and thus no sense of the stability or fragility of the trained
% algorithm.
The similarity of the training sets in CV produces dependent errors across the test sets on
different folds. In spite of the simple correlation structure of that dependence, those correlations
cannot be estimated unbiasedly under all distributions. This is a key point in the analysis by
Bengio and Grandvalet \citep{Bengio2004NoUnbiasedEstKCV}, who prove that there is thus no universal
unbiased estimator of the variance of $K$-fold CV. Several authors have attempted to derive
approximate estimators of this variance that depend on the distribution of the errors and the nature
of the learning algorithm; a review is included in the work of Markatou et
al. \citep{Markatou2005ANOVAofCVofGenError} on this problem. All of these authors acknowledge its
general difficulty when the figure of merit is an elementary error measure---without addressing the
further issues associated with ROC analysis.

\bigskip

Furthermore, it can be shown that cross validation is not a smooth estimator (small changes in the
input can lead to large changes in the output \citep{Efron1997ImprovementsOnCross, Yousef2006DSc});
this is the source of its susceptibility to high variability---often a very important practical
issue. Therefore it cannot serve as the basis for methods for estimating the variance of performance
based on the statistical influence function, a form of statistical perturbation or variational
analysis, that depends on the existence of certain derivatives
\citep[][Ch. 2]{Huber2004RobustStatistics} and \cite{Yousef2019EstimatingStandardErrorCross-arxiv,
  Yousef2009EstCVvariability}.

\bigskip

Our own research into this problem \citep{Yousef2005EstimatingThe}, builds on the work of Efron and
Tibshirani \cite{Efron1997ImprovementsOnCross}. We use the statistical influence function (or
``infinitesimal jackknife'') to estimate the uncertainty in performance assessment from both finite
training and finite testing \citep{Yousef2005EstimatingThe}. All of the available samples are
pooled, but the traditional hygiene is respected step-by-step: trainers are obtained by drawing a
bootstrap sample and the resulting model is tested only on the samples not included in that
bootstrap. This is repeated a large number of times. The overall procedure is carried out using an
estimator that is a smooth version of cross validation, so the perturbational approach to estimating
confidence intervals using this input can be applied \citep{Efron1997ImprovementsOnCross}. It yields
an estimate of the unconditional mean performance as well as an estimate of its finite-sample
uncertainty using the very same bootstrap samples.

\bigskip

This approach to estimating the unconditional mean and the variance of this estimate can be used to
assess the performance of a single classifier as well as to assess the difference in performance
across competing classifiers (W. Chen, R.F. Wagner, W.A. Yousef, \textit{loc. cit.}). The latter is
what is desired in studies whose goal is to determine whether algorithmic combinations of multiple
genomic or proteomic biomarkers add information to that already available from combining
conventional clinical biomarkers such as estrogen receptor status, age, and other more readily
accessible conditions of the patient \citep{Hess2006PharmacogenomicPre}.

\bigskip

A practical by-product of this procedure is that it offers one approach to the problem of sizing a
database for a given diagnostic task---a fundamental issue for the field of multiple biomarkers as
well as bioinformatics in general. One carries out a pilot study using the paradigm just sketched,
and obtains an estimate of the confidence intervals of interest. One can then design a pivotal study
with a target confidence interval by scaling up the size of the pilot study as described
below. While it is important to recognize that such estimates may be subject to large uncertainties
themselves from the finite sample sizes used, it is also unwise to proceed without such a guide.

\section{A PROPOSAL FOR A STRATEGY FOR ASSESSING THE MULTIPLE-BIOMARKER CLASSIFIER PROBLEM}
\begin{figure}[t]\centering
  \hfil\includegraphics[height=0.2\textheight]{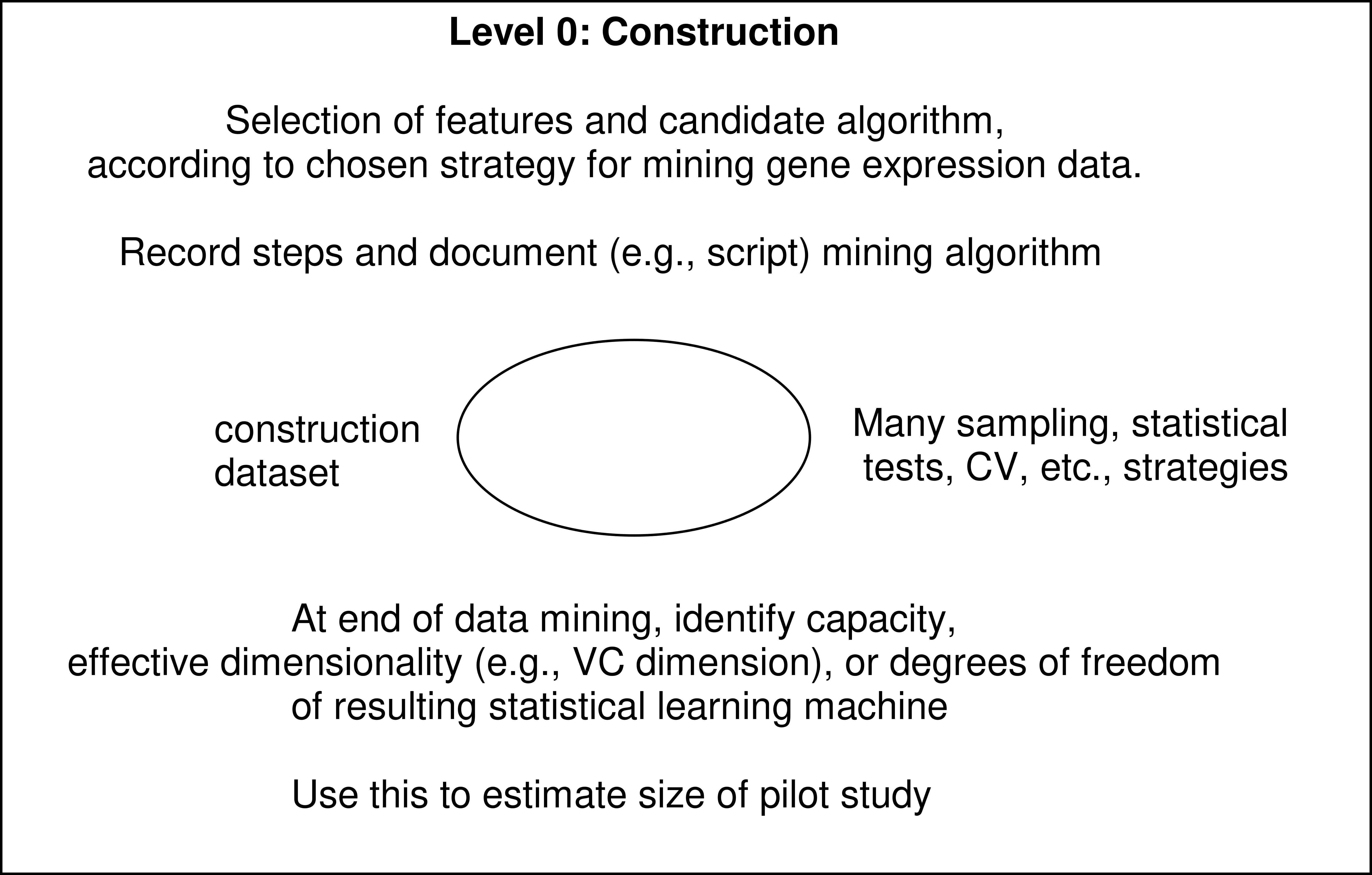}\hfil\includegraphics[height=0.2\textheight]{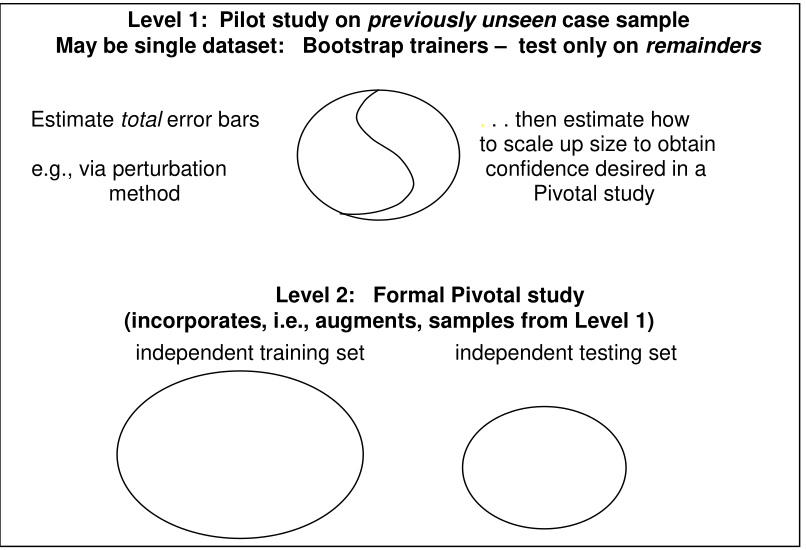}\hfil
  \caption{Schematic summary of Level 0 (the construction stage), Level 1 (the pilot study), and
    Level 2 (the pivotal study).}\label{FIGLevel}%
\end{figure}
Since it is not obvious that a given multiple-biomarker problem is well-posed (or well-conditioned)
even after the completion of the construction process, it seems reasonable to require a higher level
of overall validation than is typical in our experience. We recommend the problem be analyzed at
three levels or stages. A schematic sketch of the approach is provided in Figure \ref{FIGLevel}

\subsection{Level zero, or ground floor}
Investigators who are mining microarray data ultimately come up with their own favorite approach to
feature selection, use of prior information (or not), development of the general architecture for
their classifier, and then its training. Because of all of the new issues, it is reasonable to ask
investigators to keep and report the equivalent of a laboratory notebook on the procedure they
followed in arriving at their features and classifier. We will speak of this level as the ``ground
floor.'' Since there is already a large literature on this complex subject
\citep{Parmigiani2003AnaGene, Speed2003StatAnaGene, McLachlan2004DiscAna, Simon2003DesignAnaDNA}, we
do not address the subjects of selection of features or classifier architecture further here. Our
focus is on methods of assessment of classifiers and the implications of these methods for study
design.

\bigskip

Some guidance on the recommended sample size at this development and training stage has been
provided by Dobbin and Simon \citep{Dobbin2007SampleSizePlannig}. They have developed an approach
based on a multivariate normal model for mean or expected levels of performance, but acknowledge
that they do not control for the variance due to the finite training sample (recall that the latter
is a central point of the present paper). This may be reasonable when the class separability is in
fact large (see, e.g., \cite{GolubEtal1999MolecClass}, where it is visually obvious).

\subsection{Level one, or pilot study}\label{sec:level-one-or}
Some form of pilot study, using only previously unseen cases, is recommended next. One approach to
this pilot study is to incorporate use of the bootstrap and statistical Influence Function (IF) as
in \cite{Efron1997ImprovementsOnCross, Yousef2005EstimatingThe}. In that work, there is only one
dataset at this stage. A bootstrap sample is drawn and used as the training set for the combination
of biomarkers and classifier architecture that result from level zero; the remainders from the
dataset are used as the testers. This is repeated within a framework that allows estimation of the
mean performance as well as the variance of this estimate using the influence function. Since in the
limit of very large datasets a bootstrap sample is expected to be supported on only 0.632 of the
data \citep{Efron1983EstimatingTheError}, one might expect this procedure to be closely related to
2:1 (trainers:testers) cross validation. On modest finite datasets in the experience of
\cite{Efron1997ImprovementsOnCross, Yousef2006DSc, Yousef2004ComparisonOf} it appears to behave in
the mean more like 1:1 cross validation ($K$-fold CV, with $K=2$). However, as mentioned above and
to be recounted here, it has more generality than cross validation. It may be thought of as
generalized cross validation or a smooth version of cross validation
\citep{Efron1983EstimatingTheError, Efron1997ImprovementsOnCross}.

\bigskip

In the light of the previous discussion, how large a sample can be recommended for the pilot study?
The approach of Dobbin and Simon \citep{Dobbin2007SampleSizePlannig} may offer some guidance here,
but it has the limitation cited above and there are further issues. Since the method of generalized
CV that we advocate behaves in the mean like 1:1 CV, only very roughly half of the pilot sample size
is effective in training. Recall next that one cannot expect training results to generalize until
the training set is some multiple---depending on problem complexity, class separability, and authors
consulted---in the order of 5, 10, 20 times the effective dimensionality of the learning machine. If
we select that multiple to be five, then the factor of two from the top of this paragraph suggests
that the pilot study should be ten times the size of the effective dimensionality of the learning
algorithm. One wants to avoid the situation where the bootstrapped training iterations find many
near-perfect classifiers---purely as a result of high dimensional geometry (the theorem of Cover
\citep{Cover1965GeomStatProp})---only to find them marginal or useless in the testing set. Although
\cite{Dobbin2007SampleSizePlannig} may be a useful guide for the classes of problem considered
there, there is no general solution to this initial design question---aside from attempting to
simulate a family of problems with characteristics similar to the available data.

\bigskip

There may be several candidate methods for analyzing the pilot study data. We recommend an approach
that carries out ROC analysis and includes the influence function to estimate the total uncertainty
as in \cite{Yousef2005EstimatingThe}; this is also an efficient use of the pilot study data since it
uses the summary measure of the area under the ROC curve, which typically has much smaller variance
than estimates of sensitivity (or true-positive fraction) alone because it averages over a range of
values of sensitivity \citep{Wagner2007AssMedImgTutorial}. In any case, since there is no added
burden in measuring the entire empirical ROC curve, as opposed to a single (sensitivity,
specificity) point, we see no reason not to measure the entire ROC, and its variability in terms of
an area measure.

\bigskip

The estimated variability of the results of the pilot study can be used to estimate the size of a
pivotal study with a desired level of uncertainty. The mathematical scaling law at the population
level has been polynomial in our experience \citep{Beiden2003AGeneralModel}, not simple
linear. Earlier we mentioned the quadratic dependence of the variance in the inverse of the number
of training samples if the classifier is in the neighborhood of the Bayes' classifier. This
neighborhood may be a rare situation in practice, so the more conservative posture would be to
assume that an estimate of this contribution, and thus the total variance including the
contributions from the finite test set, does not scale any faster than linearly in the inverse of
the total number of samples. (``Interaction'' or cross-terms can scale faster
\citep{Beiden2003AGeneralModel}; thus the position we advocate here may be mildly conservative.) The
total variance desired from a pivotal study---compared to the variance observed in the pilot
study---then leads to a target size of the pivotal study using the conservative linear scaling.

\bigskip

A question that now arises naturally is this: What is the classifier that results from this pilot
study? A natural answer is that---once the mean performance and uncertainty have been estimated
using smooth or generalized cross-validation and calculation of the influence function---one can
then use the full pool of samples in the pilot dataset to train the combination of biomarkers and
learning machine architecture selected at the end of level zero and used throughout this level one.

\bigskip

Finally, the samples from the pilot study are not then replaced for the pivotal study, rather they
are augmented to achieve the total target size discussed above. In this way the pilot study may be
considered an ``interim look'' with respect to the overall pivotal study\footnote{G. Campbell and
  E. Russek-Cohen, CDRH/FDA, April 12, 2007 (personal communication with Robert F. Wagner).}.

\subsection{Level two, or pivotal study}
Many investigators assume that the most conservative approach to a pivotal study is to carry out an
``independent external validation,'' i.e., to simply obtain a test set completely independent of the
training set(s). Given a great deal of attention to the details of obtaining the dataset for
external validation, this is likely to achieve its desired goal. However, it is good to keep in mind
the exchange between Tibshirani and Dave cited earlier \citep{Tibshirani2005ImmSigLymphoma}. In that
light, one might argue for further resampling (including simple swapping) across the training and
test sets---even at the stage of external validation. In our approach the role of external
validation is split into pilot and pivotal studies. We also depend on resampling in our approach to
a pivotal study---but the complete separation of the training and test sets, what we call ``the
traditional hygiene,'' is maintained.

\bigskip

Our approach to performing a pivotal study was given in \cite{Yousef2006AssessClass}, and extended
in~\cite{Chen2012ClassVar}, for the situation where the figure of merit was the total area under
the ROC curve and methods from $U$-statistics were used to obtain nonparametric estimates of
performance. That work is readily extended to the situation where the figure of merit is the partial
area under a specified region of the ROC curve by a simple adjustment of the kernel in the
Mann-Whitney-Wilcoxon figure of merit \cite{Yousef2013PAUC}. Implementation of this approach using
bootstrap resampling to probe the uncertainty due to the finite training set leads to the limitation
to the expected 0.632 support on the trainers (or 0.5 support in many practical examples
\citep{Efron1997ImprovementsOnCross, Yousef2006DSc, Yousef2004ComparisonOf}), and is therefore
vulnerable to some conservative bias. In practice, this mean effect was dominated by variance
effects, but it is important to recognize that it is present.

\bigskip

The methods from $U$-statistics cited here could also be applied in the pilot study if investigators
are averse to working with a pooled dataset at that stage and prefer completely isolated training
and testing sets. This, however, would lead to a loss in efficiency.

\section{FURTHER PRACTICAL RECOMMENDATIONS}
The lack of reproducibility of so many studies in this general field has led reviewers to recommend
to investigators that they make their work completely transparent to their readers. A high-profile
example is that of Petricoin, Liotta and colleagues who used features from mass spectrometry to
develop a proteomic classifier for discriminating between the presence or absence of ovarian cancer
\citep{Petricoin2002UseProteomic}. The performance of their classifier was considered to have been
validated by these investigators using independent test samples. However, their results have not
been reproduced and have been subjected to challenges and much debate based on potential biases in
the way the samples were analyzed (see, e.g., \cite{Ransohoff2005BiasAsThreat}, among many). A
follow-up editorial in Nature \citep{EditorialNature2004Protemoics} offered the following wisdom:
``\ldots perhaps the most important aspect of the debate is that it would never have arisen if
Liotta and Petricoin had not posted their data on the Internet. The episode underscores the crucial
importance of readily available public data for scientific progress and, ultimately, for public
health.''

\bigskip

Hastie and Tibshirani, after a review of the issues associated with the work of S.S. Dave
et. al. \citep{Dave2004PredOfSurv}, have generated their own list of recommendations to
practitioners in this field (Course Notes: Statistical Learning and Data Mining II. Tools for Tall
and Wide Data, Philadelphia PA., October 12-13, 2006). They include:%
\begin{itemize}[partopsep=0in,parsep=0in,topsep=0.1in,itemsep=0.0in,leftmargin=0.2in]
  \item Encourage authors to publish not only the raw data, but a script of their analysis.
  \item Encourage authors to use ``canned'' methods/packages, with built-in cross validation to
  validate the model search process.
  \item Develop measures of the fragility of an analysis.
\end{itemize}
The first two of these suggestions speak for themselves. Their third suggestion is addressed by the
methods of estimating the total uncertainty of performance estimates discussed throughout the
sections on performance assessment in the present paper.

\bigskip

A critical review of published microarray studies in the field of cancer outcomes has been published
by Dupuy and Simon \citep{Dupuy2007CriticalRevMAstudies}. It includes forty guidelines and
recommendations prompted by the research of those authors and the problems they found in the
literature.

\bigskip

Finally, a perennial issue in the field of statistical learning machines is that of reuse of the
independent test set after modifications to an original designed and tested algorithm. Such a
process turns the test set into part of the design or training set
\citep{Gur2004OnTheRepeatedUse}. Ground rules must be developed for avoiding this approach and
penalizing it when it occurs. Moreover, because of the complexity of the overall problem being
addressed here, many authors even argue for the necessity of several independent external test
sets\footnote{e.g., K. Hess, Seventh MAQC Conference, SAS Institute, Cary NC, May 24-25, 2007
  (personal communication with Robert F. Wagner).}

\section{SUMMARY}
We have proposed a three-level strategy for the assessment of a multiple-biomarker classifier. The
approach uses some fundamental principles from the field of statistical learning to guide the sizing
of a pilot study; the results of the pilot study are used to guide the sizing of a pivotal
study. This approach may be slightly more elaborate than some might prefer, but at a minimum it
addresses the limitations of alternative approaches mentioned above. Moreover, at the end of the
pilot study, the investigator can consider the point estimate and the estimated error bars to see if
their present classifier development strategy is taking them in a desirable direction. If so, then
they continue on toward the pivotal study; if not, their original approach might be abandoned in
favor of another one. It is therefore possible that in the long run the overall strategy outlined
here is an efficient and thus least-burdensome use of resources.

% The authors and their colleagues have begun to carry out a demonstration project of the stages
% described here, using real clinical datasets. (W. Chen, R.F. Wagner, W.A. Yousef,
% \textit{loc. cit.}) Results of this large project are about to be published sequentially.

\section{ACKNOWLEDGMENTS}
\begin{itemize}[partopsep=0in,parsep=0in,topsep=0in,itemsep=0.0in,leftmargin=0.1in]
  \item The author is indebted to Robert F. Wagner, Senior Biomedical Research Scientist at Center
  for Devices and Radiological Health (CDRH), Food and Drug Administration (FDA), Silver Spring, MD
  20993, USA, who passed away in 2008. Robert F. Wagner (the supervisor) or Bob Wagner (the big
  brother and friend) is aware of this manuscript more than anyone else. Every word in this
  manuscript attests his motivation, ideas, support, scientific communication with different
  parties, and even his writing style that I learned from.

  \item The author gratefully acknowledges the general chair of the MAQC2 Project, Leming Shi (NCTR/FDA),
  and the chair of its Biostatistics Working Group, Gregory Campbell (CDRH/FDA), for discussions of
  this work within the project with Robert F. Wagner and inviting its presentation at their May 24-25
  2007 Conference at the SAS Institute in Cary NC and follow-up Conference March 24-26 2008 at the
  Food and Drug Administration in Rockville MD.

  \item The author also thanks Weijie Chen (CDRH/FDA), Tim Davison (Asuragen), Kenneth Hess (MD Anderson),
  Russ Wolfinger (SAS), and Bill Worzel (Genetics Squared) of the same consortium for helpful
  suggestions and for facilitating communications with Robert F. Wagner.

  \item Finally, this work was supported in part by a CDRH Medical Device Fellowship awarded to Waleed
  A. Yousef and administered by the Oak Ridge Institute for Science and Education (ORISE).
\end{itemize}

\section{DISCLAIMER}
This paper represents the professional views of the author. It is not an official document,
guidance, or policy of the US government, Department of Health and Human Services, or the Food and
Drug Administration, nor should any official endorsement be inferred. Similarly, it is not an
official document of the Microarray Quality Control-Phase 2 (MAQC2) project, nor should its
endorsement be inferred.

\bibliographystyle{elsarticle-num}
\bibliography{booksIhave,publications}

\begin{thebibliography}{10}
\expandafter\ifx\csname url\endcsname\relax
  \def\url#1{\texttt{#1}}\fi
\expandafter\ifx\csname urlprefix\endcsname\relax\def\urlprefix{URL }\fi
\expandafter\ifx\csname href\endcsname\relax
  \def\href#1#2{#2} \def\path#1{#1}\fi

\bibitem{Shi2010MAQCII}
L.~Shi, G.~Campbell, W.~D. Jones, F.~Campagne, Z.~Wen, S.~J. Walker, Z.~Su,
  T.~M. Chu, F.~M. Goodsaid, L.~Pusztai, J.~D. {Shaughnessy Jr.}, A.~Oberthuer,
  R.~S. Thomas, R.~S. Paules, M.~Fielden, B.~Barlogie, W.~Chen, P.~Du,
  M.~Fischer, C.~Furlanello, B.~D. Gallas, X.~Ge, D.~B. Megherbi, W.~F.
  Symmans, M.~D. Wang, J.~Zhang, H.~Bitter, B.~Brors, P.~R. Bushel, M.~Bylesjo,
  M.~Chen, J.~Cheng, J.~Chou, T.~S. Davison, M.~Delorenzi, Y.~Deng,
  V.~Devanarayan, D.~J. Dix, J.~Dopazo, K.~C. Dorff, F.~Elloumi, J.~Fan,
  S.~Fan, X.~Fan, H.~Fang, N.~Gonzaludo, K.~R. Hess, H.~Hong, J.~Huan, R.~A.
  Irizarry, R.~Judson, D.~Juraeva, S.~Lababidi, C.~G. Lambert, L.~Li, Y.~Li,
  Z.~Li, S.~M. Lin, G.~Liu, E.~K. Lobenhofer, J.~Luo, W.~Luo, M.~N. McCall,
  Y.~Nikolsky, G.~A. Pennello, R.~G. Perkins, R.~Philip, V.~Popovici, N.~D.
  Price, F.~Qian, A.~Scherer, T.~Shi, W.~Shi, J.~Sung, D.~Thierry-Mieg,
  J.~Thierry-Mieg, V.~Thodima, J.~Trygg, L.~Vishnuvajjala, S.~J. Wang, J.~Wu,
  Y.~Wu, Q.~Xie, W.~A. Yousef, L.~Zhang, X.~Zhang, S.~Zhong, Y.~Zhou, S.~Zhu,
  D.~Arasappan, W.~Bao, A.~B. Lucas, F.~Berthold, R.~J. Brennan, A.~Buness,
  J.~G. Catalano, C.~Chang, R.~Chen, Y.~Cheng, J.~Cui, W.~Czika, F.~Demichelis,
  X.~Deng, D.~Dosymbekov, R.~Eils, Y.~Feng, J.~Fostel, S.~Fulmer-Smentek, J.~C.
  Fuscoe, L.~Gatto, W.~Ge, D.~R. Goldstein, L.~Guo, D.~N. Halbert, J.~Han,
  S.~C. Harris, C.~Hatzis, D.~Herman, J.~Huang, R.~V. Jensen, R.~Jiang, C.~D.
  Johnson, G.~Jurman, Y.~Kahlert, S.~A. Khuder, M.~Kohl, J.~Li, M.~Li, Q.~Z.
  Li, S.~Li, J.~Liu, Y.~Liu, Z.~Liu, L.~Meng, M.~Madera, F.~Martinez-Murillo,
  I.~Medina, J.~Meehan, K.~Miclaus, R.~A. Moffitt, D.~Montaner, P.~Mukherjee,
  G.~J. Mulligan, P.~Neville, T.~Nikolskaya, B.~Ning, G.~P. Page, J.~Parker,
  R.~M. Parry, X.~Peng, R.~L. Peterson, J.~H. Phan, B.~Quanz, Y.~Ren,
  S.~Riccadonna, A.~H. Roter, F.~W. Samuelson, M.~M. Schumacher, J.~D.
  Shambaugh, Q.~Shi, R.~Shippy, S.~Si, A.~Smalter, C.~Sotiriou, M.~Soukup,
  F.~Staedtler, G.~Steiner, T.~H. Stokes, Q.~Sun, P.~Y. Tan, R.~Tang, Z.~Tezak,
  B.~Thorn, M.~Tsyganova, Y.~Turpaz, S.~C. Vega, R.~Visintainer, J.~von Frese,
  C.~Wang, E.~Wang, J.~Wang, W.~Wang, F.~Westermann, J.~C. Willey, M.~Woods,
  S.~Wu, N.~Xiao, J.~Xu, L.~Xu, L.~Yang, X.~Zeng, M.~Zhang, C.~Zhao, R.~K.
  Puri, U.~Scherf, W.~Tong, R.~D. Wolfinger, The microarray quality control
  (maqc)-ii study of common practices for the development and validation of
  microarray-based predictive models, Nat Biotechnol 28~(8) (2010) 827--838
  (2010).

\bibitem{Butcher2004SysBioDrug}
E.~C. Butcher, E.~L. Berg, E.~J. Kunkel, \href{https://doi.org/nbt1017 [pii]
  10.1038/nbt1017}{{Systems Biology in Drug discovery}}, Nat Biotechnol 22~(10)
  (2004) 1253--1259 (2004).
\newblock \href {https://doi.org/nbt1017 [pii] 10.1038/nbt1017}
  {\path{doi:nbt1017 [pii] 10.1038/nbt1017}}.
\newline\urlprefix\url{https://doi.org/nbt1017 [pii] 10.1038/nbt1017}

\bibitem{Shi2006MAQCproj}
L.~Shi, L.~H. Reid, W.~D. Jones, R.~Shippy, J.~A. Warrington, S.~C. Baker,
  P.~J. Collins, F.~de~Longueville, E.~S. Kawasaki, K.~Y. Lee, Y.~Luo, Y.~A.
  Sun, J.~C. Willey, R.~A. Setterquist, G.~M. Fischer, W.~Tong, Y.~P. Dragan,
  D.~J. Dix, F.~W. Frueh, F.~M. Goodsaid, D.~Herman, R.~V. Jensen, C.~D.
  Johnson, E.~K. Lobenhofer, R.~K. Puri, U.~Schrf, J.~Thierry-Mieg, C.~Wang,
  M.~Wilson, P.~K. Wolber, L.~Zhang, S.~Amur, W.~Bao, C.~C. Barbacioru, A.~B.
  Lucas, V.~Bertholet, C.~Boysen, B.~Bromley, D.~Brown, A.~Brunner, R.~Canales,
  X.~M. Cao, T.~A. Cebula, J.~J. Chen, J.~Cheng, T.~M. Chu, E.~Chudin,
  J.~Corson, J.~C. Corton, L.~J. Croner, C.~Davies, T.~S. Davison,
  G.~Delenstarr, X.~Deng, D.~Dorris, A.~C. Eklund, X.~H. Fan, H.~Fang,
  S.~Fulmer-Smentek, J.~C. Fuscoe, K.~Gallagher, W.~Ge, L.~Guo, X.~Guo,
  J.~Hager, P.~K. Haje, J.~Han, T.~Han, H.~C. Harbottle, S.~C. Harris,
  E.~Hatchwell, C.~A. Hauser, S.~Hester, H.~Hong, P.~Hurban, S.~A. Jackson,
  H.~Ji, C.~R. Knight, W.~P. Kuo, J.~E. LeClerc, S.~Levy, Q.~Z. Li, C.~Liu,
  Y.~Liu, M.~J. Lombardi, Y.~Ma, S.~R. Magnuson, B.~Maqsodi, T.~McDaniel,
  N.~Mei, O.~Myklebost, B.~Ning, N.~Novoradovskaya, M.~S. Orr, T.~W. Osborn,
  A.~Papallo, T.~A. Patterson, R.~G. Perkins, E.~H. Peters, R.~Peterson, K.~L.
  Philips, P.~S. Pine, L.~Pusztai, F.~Qian, H.~Ren, M.~Rosen, B.~A. Rosenzweig,
  R.~R. Samaha, M.~Schena, G.~P. Schroth, S.~Shchegrova, D.~D. Smith,
  F.~Staedtler, Z.~Su, H.~Sun, Z.~Szallasi, Z.~Tezak, D.~Thierry-Mieg, K.~L.
  Thompson, I.~Tikhonova, Y.~Turpaz, B.~Vallanat, C.~Van, S.~J. Walker, S.~J.
  Wang, Y.~Wang, R.~Wolfinger, A.~Wong, J.~Wu, C.~Xiao, Q.~Xie, J.~Xu, W.~Yang,
  S.~Zhong, Y.~Zong, W.~{Slikker Jr.}, \href{https://doi.org/nbt1239 [pii]
  10.1038/nbt1239}{{The Microarray Quality Control (maqc) Project Shows Inter-
  and Intraplatform Reproducibility of Gene Expression measurements}}, Nat
  Biotechnol 24~(9) (2006) 1151--1161 (2006).
\newblock \href {https://doi.org/nbt1239 [pii] 10.1038/nbt1239}
  {\path{doi:nbt1239 [pii] 10.1038/nbt1239}}.
\newline\urlprefix\url{https://doi.org/nbt1239 [pii] 10.1038/nbt1239}

\bibitem{Klebanov2007StatMethodMicroArr}
L.~Klebanov, X.~Qiu, S.~Welle, A.~Yakovlev, \href{https://doi.org/nbt0107-25
  [pii] 10.1038/nbt0107-25}{{Statistical Methods and Microarray data}}, Nat
  Biotechnol 25~(1) (2007) 25--27 (2007).
\newblock \href {https://doi.org/nbt0107-25 [pii] 10.1038/nbt0107-25}
  {\path{doi:nbt0107-25 [pii] 10.1038/nbt0107-25}}.
\newline\urlprefix\url{https://doi.org/nbt0107-25 [pii] 10.1038/nbt0107-25}

\bibitem{Liang2007MAQCpapersCracks}
P.~Liang, \href{https://doi.org/nbt0107-27 [pii] 10.1038/nbt0107-27}{{MAQC
  Papers Over the cracks}}, Nat Biotechnol 25~(1) (2007) 27--29 (2007).
\newblock \href {https://doi.org/nbt0107-27 [pii] 10.1038/nbt0107-27}
  {\path{doi:nbt0107-27 [pii] 10.1038/nbt0107-27}}.
\newline\urlprefix\url{https://doi.org/nbt0107-27 [pii] 10.1038/nbt0107-27}

\bibitem{Chen2012UncertEst}
W.~Chen, W.~A. Yousef, B.~D. Gallas, E.~R. Hsu, S.~Lababidi, R.~Tang, G.~A.
  Pennello, W.~F. Symmans, L.~Pusztai, {Uncertainty Estimation With a Finite
  Dataset in the Assessment of Classification models}, Computational Statistics
  {\&} Data Analysis 56~(5) (2012) 1016--1027 (2012).
\newblock \href {https://doi.org/10.1016/j.csda.2011.05.024}
  {\path{doi:10.1016/j.csda.2011.05.024}}.

\bibitem{Wagner2008FiniteTrainingRadSLM}
R.~F. Wagner, W.~A. Yousef, W.~Chen, {Finite Training of Radiologists and
  Statistical Learning Machines: Parallel Lessons}, in: A.~B. Wolbarst, K.~L.
  Mossman, W.~R. Hendee (Eds.), Advances in Medical Physics: 2008, Medical
  Physics Pub., Madison, WI, 2008 (2008).

\bibitem{Parmigiani2003AnaGene}
G.~Parmigiani, {The analysis of gene expression data : methods and software},
  Springer-Verlag, New York, 2003 (2003).

\bibitem{Speed2003StatAnaGene}
T.~P. Speed, {Statistical analysis of gene expression microarray data}, Chapman
  $\backslash${\&} Hall/CRC, Boca Raton, FL, 2003 (2003).

\bibitem{McLachlan2004DiscAna}
G.~J. McLachlan,
  \href{http://www.loc.gov/catdir/enhancements/fy0626/2005271842-b.html
  http://www.loc.gov/catdir/enhancements/fy0626/2005271842-d.html
  http://www.loc.gov/catdir/enhancements/fy0626/2005271842-t.html}{{Discriminant
  analysis and statistical pattern recognition}}, Hoboken, N.J., 2004 (2004).
\newline\urlprefix\url{http://www.loc.gov/catdir/enhancements/fy0626/2005271842-b.html
  http://www.loc.gov/catdir/enhancements/fy0626/2005271842-d.html
  http://www.loc.gov/catdir/enhancements/fy0626/2005271842-t.html}

\bibitem{Simon2003DesignAnaDNA}
R.~M. Simon, E.~L. Korn, L.~M. McShane, M.~D. Radmacher, G.~W. Wright, Y.~Zhao,
  {Design and analysis of DNA microarray investigations}, Springer, New York,
  2003 (2003).

\bibitem{Hastie2001TheElements}
T.~Hastie, R.~Tibshirani, J.~H. Friedman, {The elements of statistical learning
  : data mining, inference, and prediction}, Springer, New York, 2001 (2001).

\bibitem{Yousef2008SLM-arxiv}
W.~A. Yousef, A review of statistical learning machines from atr to dna
  microarrays: design, assessment, and advice for practitioners, arXiv preprint
  arXiv:1906.10019 (2019).

\bibitem{Cover1965GeomStatProp}
T.~M. Cover, {Geometrical and Statistical Properties of Systems of Linear
  Inequalities With Applications in Pattern recognition}, IEEE Transactions on
  Electronic Computers (1965) 326--334 (1965).

\bibitem{Baum1989WhatSize}
E.~B. Baum, H.~D., {What Size Net Gives Valid generalization?}, Neural
  Computation 1~(1) (1989) 151--160 (1989).

\bibitem{Vapnik2000Nature}
V.~N. Vapnik, {The nature of statistical learning theory}, 2nd Edition,
  Springer, New York, 2000 (2000).

\bibitem{Fukunaga1990Introduction}
K.~Fukunaga, {Introduction to statistical pattern recognition}, 2nd Edition,
  Academic Press, Boston, 1990 (1990).

\bibitem{Fukunaga1989EstimationOf}
K.~Fukunaga, R.~R. Hayes, {Estimation of Classifier performance}, Pattern
  Analysis and Machine Intelligence, IEEE Transactions on - 11~(- 10) (1989) --
  1101 (1989).

\bibitem{Yousef2019PrudenceWhenAssumingNormality-arxiv}
W.~A. Yousef, Prudence when assuming normality: an advice for machine learning
  practitioners, arXiv preprint arXiv:1907.12852 (2019).

\bibitem{Efron1983EstimatingTheError}
B.~Efron, {Estimating the Error Rate of a Prediction Rule: Improvement on
  Cross-Validation}, Journal of the American Statistical Association 78~(382)
  (1983) 316--331 (1983).

\bibitem{Efron1997ImprovementsOnCross}
B.~Efron, R.~Tibshirani, {Improvements on Cross-Validation: the $.632+$
  Bootstrap Method}, Journal of the American Statistical Association 92~(438)
  (1997) 548--560 (1997).

\bibitem{Dobbin2007SampleSizePlannig}
K.~K. Dobbin, R.~M. Simon, {Sample Size Planning for Developing Classifiers
  Using High-Dimensional Dna Microarray Data}, Biostatistics 8~(1) (2007)
  101--117 (2007).

\bibitem{Walter2005ThePartialArea}
S.~D. Walter, {The Partial Area Under the Summary {\{}ROC{\}} curve},
  Statistics In Medicine 24~(13) (2005) 2025--2040 (2005).

\bibitem{Jiang1996AReceiverOperating}
Y.~Jiang, C.~E. Metz, R.~M. Nishikawa, {A Receiver Operating Characteristic
  Partial Area Index for Highly Sensitive Diagnostic tests}, Radiology 201~(3)
  (1996) 745--750 (1996).

\bibitem{Dodd2003PAUCEstReg}
L.~E. Dodd, M.~S. Pepe, {Partial {AUC} Estimation and Regression}, Biometrics
  59~(3) (2003) 614--623 (2003).

\bibitem{Yousef2013PAUC}
W.~A. Yousef, \href{https://doi.org/10.1016/j.csda.2013.02.032}{{Assessing
  Classifiers in Terms of the Partial Area Under the Roc curve}}, Computational
  Statistics {\&} Data Analysis 64~(0) (2013) 51--70 (2013).
\newline\urlprefix\url{https://doi.org/10.1016/j.csda.2013.02.032}

\bibitem{Wagner2007AssMedImgTutorial}
R.~F. Wagner, C.~E. Metz, G.~Campbell, {Assessment of Medical Imaging Systems
  and Computer Aids: a Tutorial Review}, Academic Radiology 14 (2007) 723--748
  (2007).

\bibitem{Beiden2003AGeneralModel}
S.~V. Beiden, M.~A. Maloof, R.~F. Wagner, {A General Model for Finite-Sample
  Effects in Training and Testing of Competing classifiers}, Pattern Analysis
  and Machine Intelligence, IEEE Transactions on 25~(12) (2003) 1569 (2003).

\bibitem{Chan1999ClassifierDesign}
H.~P. Chan, B.~Sahiner, R.~F. Wagner, N.~Petrick, {Classifier Design for
  Computer-Aided Diagnosis: Effects of Finite Sample Size on the Mean
  Performance of Classical and Neural Network classifiers}, Medical Physics
  26~(12) (1999) 2654--2668 (1999).

\bibitem{Yousef2006DSc}
W.~A. Yousef, M.~H. Loew, R.~F. Wagner,
  \href{http://proquest.umi.com/pqdweb?did=1083540501{\&}Fmt=7{\&}clientId=93083{\&}RQT=309{\&}VName=PQD}{{Assessment
  of statistical classification rules: implications for computational
  intelligence}}, Ph.D. thesis, Washington DC. (2006).
\newline\urlprefix\url{http://proquest.umi.com/pqdweb?did=1083540501{\&}Fmt=7{\&}clientId=93083{\&}RQT=309{\&}VName=PQD}

\bibitem{Ripley1996PRandNN}
B.~D. Ripley, {Pattern recognition and neural networks}, Cambridge University
  Press, Cambridge ; New York, 1996 (1996).

\bibitem{Duda2001PatternClassification}
R.~O. Duda, P.~E. Hart, D.~G. Stork, {Pattern classification}, 2nd Edition,
  Wiley, New York, 2001 (2001).

\bibitem{Bishop1995NeurNet}
C.~M. Bishop, {Neural networks for pattern recognition}, Clarendon Press;
  Oxford University Press., Oxford; New York, 1995 (1995).

\bibitem{Vapnik1998StatLerningTh}
V.~N. Vapnik, {Statistical learning theory}, Wiley, New York, 1998 (1998).

\bibitem{GolubEtal1999MolecClass}
T.~R. Golub, D.~K. Slonim, P.~Tamayo, C.~Huard, M.~Gaasenbeek, J.~P. Mesirov,
  H.~Coller, M.~L. Loh, J.~R. Downing, M.~A. Caligiuri, C.~D. Bloomfield, E.~S.
  Lander, {Molecular Classification of Cancer: Class Discovery and Class
  Prediction By Gene Expression Monitoring}, Science 286~(5439) (1999) 531--537
  (1999).

\bibitem{Zhang1995AssessingPrediction}
P.~Zhang, {Assessing Prediction Error in Nonparametric Regression},
  Scandinavian Journal Of Statistics 22~(1) (1995) 83--94 (1995).

\bibitem{Hastie2009ElemStat}
T.~Hastie, R.~Tibshirani, J.~H. Friedman, {The elements of statistical
  learning: data mining, inference, and prediction}, 2nd Edition, Springer, New
  York, 2009 (2009).

\bibitem{Yousef2019AUCSmoothness-arxiv}
W.~A. Yousef, {AUC}: nonparametric estiamtors and their smoothness, arXiv
  preprint arXiv:1907.12851 (2019).

\bibitem{Yousef2019LeisurelyLookVersionsVariants-arxiv}
W.~A. Yousef, A leisurely look at versions and variants of the cross validation
  estimator, arXiv preprint arXiv:1907.13413 (2019).

\bibitem{Dave2004PredOfSurv}
S.~S. Dave, G.~Wright, B.~Tan, A.~Rosenwald, R.~D. Gascoyne, W.~C. Chan, R.~I.
  Fisher, R.~M. Braziel, L.~M. Rimsza, T.~M. Grogan, T.~P. Miller, M.~LeBlanc,
  T.~C. Greiner, D.~D. Weisenburger, J.~C. Lynch, J.~Vose, J.~O. Armitage,
  E.~B. Smeland, S.~Kvaloy, H.~Holte, J.~Delabie, J.~M. Connors, P.~M.
  Lansdorp, Q.~Ouyang, T.~A. Lister, A.~J. Davies, A.~J. Norton, H.~K.
  Muller-Hermelink, G.~Ott, E.~Campo, E.~Montserrat, W.~H. Wilson, E.~S. Jaffe,
  R.~Simon, L.~Yang, J.~Powell, H.~Zhao, N.~Goldschmidt, M.~Chiorazzi, L.~M.
  Staudt, {Prediction of Survival in Follicular Lymphoma Based on Molecular
  Features of Tumor-Infiltrating Immune cells}, New England Journal of Medicine
  November 351~(21) (2004) 2159--2169 (2004).

\bibitem{Tibshirani2005ImmSigLymphoma}
R.~Tibshirani, \href{https://doi.org/352/14/1496 [pii]
  10.1056/NEJM200504073521422}{{Immune Signatures in Follicular lymphoma}}, N
  Engl J Med 352~(14) (2005) 1496--1497 (2005).
\newblock \href {https://doi.org/352/14/1496 [pii] 10.1056/NEJM200504073521422}
  {\path{doi:352/14/1496 [pii] 10.1056/NEJM200504073521422}}.
\newline\urlprefix\url{https://doi.org/352/14/1496 [pii]
  10.1056/NEJM200504073521422}

\bibitem{Yousef2006AssessClass}
W.~A. Yousef, R.~F. Wagner, M.~H. Loew, {Assessing Classifiers From Two
  Independent Data Sets Using {\{}ROC{\}} Analysis: a Nonparametric Approach},
  Pattern Analysis and Machine Intelligence, IEEE Transactions on 28~(11)
  (2006) 1809--1817 (2006).

\bibitem{Chen2012ClassVar}
W.~Chen, B.~D. Gallas, W.~A. Yousef,
  \href{https://doi.org/10.1016/j.patcog.2011.12.024}{{Classifier Variability:
  Accounting for Training and testing}}, Pattern Recognition 45~(7) (2012)
  2661--2671 (2012).
\newblock \href {https://doi.org/10.1016/j.patcog.2011.12.024}
  {\path{doi:10.1016/j.patcog.2011.12.024}}.
\newline\urlprefix\url{https://doi.org/10.1016/j.patcog.2011.12.024}

\bibitem{Bengio2004NoUnbiasedEstKCV}
Y.~Bengio, Y.~Grandvalet, {No Unbiased Estimator of the Variance of K-Fold
  Cross-Validation}, J. Mach. Learn. Res. 5 (2004) 1089--1105 (2004).

\bibitem{Markatou2005ANOVAofCVofGenError}
M.~Markatou, H.~Tian, S.~Biswas, G.~Hripcsak, {Analysis of Variance of
  Cross-Validation Estimators of the Generalization Error}, J. Mach. Learn.
  Res. 6 (2005) 1127--1168 (2005).

\bibitem{Huber2004RobustStatistics}
P.~J. Huber, {Robust statistics}, Wiley-Interscience, Hoboken, N.J., 2004
  (2004).

\bibitem{Yousef2019EstimatingStandardErrorCross-arxiv}
W.~A. Yousef, Estimating the standard error of cross-validation-based
  estimators of classification rules performance, arXiv preprint
  arXiv:1908.00325 (2019).

\bibitem{Yousef2009EstCVvariability}
W.~A. Yousef, W.~Chen, {Estimating Cross-Validation Variability}, in:
  Proceedings of the 2009 Joint Statistical Meeting, Section on Statistics in
  Epidemiology., 2009, pp. 3318--3326 (2009).

\bibitem{Yousef2005EstimatingThe}
W.~A. Yousef, R.~F. Wagner, M.~H. Loew, {Estimating the Uncertainty in the
  Estimated Mean Area Under the {\{}ROC{\}} Curve of a Classifier}, Pattern
  Recognition Letters 26~(16) (2005) 2600--2610 (2005).

\bibitem{Hess2006PharmacogenomicPre}
K.~R. Hess, K.~Anderson, W.~F. Symmans, V.~Valero, N.~Ibrahim, J.~A. Mejia,
  D.~Booser, R.~L. Theriault, A.~U. Buzdar, P.~J. Dempsey, R.~Rouzier,
  N.~Sneige, J.~S. Ross, T.~Vidaurre, H.~L. Gomez, G.~N. Hortobagyi,
  L.~Pusztai, \href{https://doi.org/10.1200/jco.2006.05.6861}{{Pharmacogenomic
  Predictor of Sensitivity To Preoperative Chemotherapy With Paclitaxel and
  Fluorouracil, Doxorubicin, and Cyclophosphamide in Breast Cancer}}, J Clin
  Oncol 24~(26) (2006) 4236--4244 (2006).
\newblock \href {https://doi.org/10.1200/jco.2006.05.6861}
  {\path{doi:10.1200/jco.2006.05.6861}}.
\newline\urlprefix\url{https://doi.org/10.1200/jco.2006.05.6861}

\bibitem{Yousef2004ComparisonOf}
W.~A. Yousef, R.~F. Wagner, M.~H. Loew, {Comparison of Non-Parametric Methods
  for Assessing Classifier Performance in Terms of {\{}ROC{\}} Parameters}, in:
  Applied Imagery Pattern Recognition Workshop, 2004. Proceedings. 33rd; IEEE
  Computer Society, 2004, pp. 190--195 (2004).

\bibitem{Petricoin2002UseProteomic}
E.~F. Petricoin, A.~M. Ardekani, B.~A. Hitt, P.~J. Levine, V.~A. Fusaro, S.~M.
  Steinberg, G.~B. Mills, C.~Simone, D.~A. Fishman, E.~C. Kohn, L.~A. Liotta,
  \href{https://doi.org/S0140-6736(02)07746-2 [pii]
  10.1016/S0140-6736(02)07746-2}{{Use of Proteomic Patterns in Serum To
  Identify Ovarian cancer}}, Lancet 359~(9306) (2002) 572--577 (2002).
\newblock \href {https://doi.org/S0140-6736(02)07746-2 [pii]
  10.1016/S0140-6736(02)07746-2} {\path{doi:S0140-6736(02)07746-2 [pii]
  10.1016/S0140-6736(02)07746-2}}.
\newline\urlprefix\url{https://doi.org/S0140-6736(02)07746-2 [pii]
  10.1016/S0140-6736(02)07746-2}

\bibitem{Ransohoff2005BiasAsThreat}
D.~F. Ransohoff, {Bias As a Threat To the Validity of Cancer Molecular-Marker
  research}, Nature Reviews Cancer 5 (2005).

\bibitem{EditorialNature2004Protemoics}
Editorial, \href{https://doi.org/10.1038/429487a 429487a [pii]}{{Proteomic
  Diagnostics tested}}, Nature 429~(6991) (2004) 487 (2004).
\newblock \href {https://doi.org/10.1038/429487a 429487a [pii]}
  {\path{doi:10.1038/429487a 429487a [pii]}}.
\newline\urlprefix\url{https://doi.org/10.1038/429487a 429487a [pii]}

\bibitem{Dupuy2007CriticalRevMAstudies}
A.~Dupuy, R.~M. Simon, \href{https://doi.org/99/2/147 [pii]
  10.1093/jnci/djk018}{{Critical Review of Published Microarray Studies for
  Cancer Outcome and Guidelines on Statistical Analysis and reporting}}, J Natl
  Cancer Inst 99~(2) (2007) 147--157 (2007).
\newblock \href {https://doi.org/99/2/147 [pii] 10.1093/jnci/djk018}
  {\path{doi:99/2/147 [pii] 10.1093/jnci/djk018}}.
\newline\urlprefix\url{https://doi.org/99/2/147 [pii] 10.1093/jnci/djk018}

\bibitem{Gur2004OnTheRepeatedUse}
D.~Gur, R.~F. Wagner, H.~P. Chan, {On the Repeated Use of Databases for Testing
  Incremental Improvement of Computer-Aided Detection schemes}, Acad Radiol
  11~(1) (2004) 103--105 (2004).

\end{thebibliography}

\end{document}